\documentclass[letterpaper, twocolumn, superscriptaddress, showpacs, preprintnumbers, amsmath, amssymb,longbibliography, nofootinbib]{revtex4-1}

\usepackage[pdftex]{graphicx}
\usepackage{caption}
\usepackage{subcaption}
\usepackage{dcolumn}
\usepackage{bm}
\usepackage{color}
\usepackage{hyperref}
\usepackage{adjustbox}
\usepackage{color}
\usepackage{float}
\usepackage{framed}
\usepackage{esvect}

\def\be{\begin{equation}}
\def\ee{\end{equation}}
\def\bea{\begin{eqnarray}}
\def\eea{\end{eqnarray}}

\newcommand{\lp}{\left(}
\newcommand{\rp}{\right)}

\begin{document}

\title{
Topological Order from Disorder and the Quantized Hall Thermal Metal: Possible Applications to the $\nu = 5/2$ State}  

\author{Chong Wang}

\author{Ashvin Vishwanath}

\author{Bertrand I. Halperin}

\affiliation{Department of Physics, Harvard University, Cambridge MA 02138, USA}

\begin{abstract}

Although numerical studies modeling the quantum hall effect at filling fraction $5/2$ predict either the Pfaffian (Pf) or its particle hole conjugate, the anti-Pfaffian (aPf) state, recent experiments appear to favor a quantized thermal hall conductivity with quantum number  $K=5/2$ , rather than the value $K=7/2$ or $K=3/2$ expected for the Pf or aPF state, respectively.     While a particle hole symmetric topological order (the PH-Pfaffian) would be consistent with the experiments, this state is believed to be energetically unfavorable in a homogenous system.   Here we study the effects of disorder that are assumed to locally nucleate domains of Pf and aPf. 
When the disorder is relatively weak and  the size of domains is relatively large, we find that when the electrical Hall conductance is on the quantized plateau with $\sigma_{xy} = (5/2)(e^2/h)$,   the value of $K$ can be only 7/2 or 3/2, with a possible first-order-like transition between them as the magnetic field is varied.  
 However, for sufficiently strong disorder an intermediate state might appear, which we analyze within a network model of the domain walls. Predominantly, we find a thermal metal phase,  where $K$ varies continuously and the longitudinal thermal conductivity is non-zero, while the  electrical Hall conductivity remains  quantized at  $(5/2) e^2/h$. However, in a  restricted parameter range we find a thermal insulator with  $K=5/2$, a disorder stabilized phase which is adiabatically connected to the PH-Pfaffian. We discuss a possible scenario to rationalize these special values of parameters.
 
\end{abstract}

\maketitle

\today

% \tableofcontents

\section{Introduction}
\label{sec:intro}

An even-denominator fractional quantized Hall (FQH) state at filling fraction $\nu=5/2$,  in a strongly-confined  two-dimensional electron system,  was first observed  by Willett et al. in 1987 \cite{Willett5/2}.  There have been debates about the nature of this state ever since. Exact diagonalization calculations in finite systems as well as density-matrix-renormalization-group calculations have strongly suggested\cite{MorfED, Feiguinetal, WangShengHaldane, StorniMorfDasSarma, RezayiSimon, PapicHaldaneRezayi, Zaleteletal, RezayiPRL} 
that the ground  state at $\nu=5/2$  should either be a state with the quantum numbers of the Pfaffian (Pf) state suggested by Moore and Read\cite{MooreMR91}, or its particle-hole conjugate, commonly denoted the anti-Pfaffian (aPf) state\cite{LevinHalperin, ssletalapf}.  The Pf and aPf states can both be thought of as paired states\cite{readgrn} of composite fermions\cite{jaincf,lopez91,hlr}, and have many common properties, including the existence of charge $e/4$ quasiparticles with non-Abelian statistics, but they have been shown to be topologically distinct.  In particular, the quantized thermal Hall conductance resulting from edge modes in the two states are predicted to have different values.
Specifically, it is predicted that if the thermal Hall conductance is written as $\kappa_{xy} = K \kappa_0 T$, where $\kappa_0 \equiv (\pi^2 k_B^2/3h)$, then   
$K$ is the chiral central charge of the boundary conformal field theory\cite{KaneFisher}, with $K = (7/2)$ for the Pf state and $K = (3/2)$ for the aPf state. 

In the limit where the participating electrons are perfectly confined to the second Landau level and are completely spin polarized, and interact with purely two-body interactions, the Pf and aPf states must have identical energies, by the particle-hole (PH) symmetry of the projected Hamiltonian. Although this degeneracy will generally be broken by Landau level mixing effects, it appears from various calculations that the energy difference at $\nu=5/2$ between the Pf and aPf states is relatively small for Hamiltonians  relevant to the experimental situation in GaAs, and there has not been unanimous agreement about which of the two states should have  lower energy (current numerical calculations seem to favor aPf\cite{RezayiSimon, Zaleteletal, RezayiPRL}).  

If one ignores effects of Landau level mixing, the choice between Pf and aPf in a finite sample could be determined by boundary effects, which clearly will violate PH symmetry.  PH symmetry will also be broken by any deviations of the electron density from the value corresponding to $\nu = 5/2$.  In particular, in either one of the two states,  the energy to create a positively charged quasiparticle, with charge $e/4$ will be different from the energy to produce a negatively charged quasiparticle, with charge $-e/4$.  If, for the sake of argument, the positive quasiparticle has the lower energy
 in the aPf state, then the negative quasiparticle will have the lower energy, by an equal amount, in the aPf state. Then in a sample with filling fraction $\nu$ slightly less than 5/2, but still within the $\nu=5/2$ quantized Hall plateau, we would  expect the system to prefer the aPf state,  whereas for $\nu$ slightly greater than 5/2, the Pf state would have lower total energy.  If the effects of Landau level mixing are not zero but are sufficiently small, the transition point between the two states might be shifted slightly away from $\nu=5/2$, but in the absence of boundary effects or of possible complications due to disorder, we would still expect a sharp transition between the two states.  If the effects of Landau level mixing are too large, then only one of the two states would exist within the allowed range of $\nu$.  Thus, if one were to measure the thermal Hall conductance at any fixed filling fraction inside the 5/2 plateau, one should obtain either $K = (3/2) $ or $K = (7/2) $, with perhaps a sharp transition between the two values as one varies the filling factor. 

 Recent measurements\cite{Mitalietal} by Banerjee et al. of the thermal conductance at $\nu=5/2$ are in dramatic disagreement with the above scenario. They obtain the result $K = (5/2) $, with a small uncertainty, which clearly distinguishes the result from ether $K = (7/2) $ or $K = (3/2) $.   The puzzle is how to explain this discrepancy.
 
Recently, the composite fermion description of the half filled Landau level \cite{hlr} was augmented to incorporate particle-hole symmetry \cite{sonphcfl}. Effectively, this theory endows  composite fermions with a Dirac character, or more precisely maps the problem to that of the Dirac surface states of three dimensional topological insulators \cite{sonphcfl,MetlitskiAV, dualdrcwts2015,WS16,geraedtsnum}. A natural extension of this inquiry is if there is a particle hole symmetric quantized Hall state for the half filled Landau level (unlike the Pfaffian and anti-Pfaffian states which are mapped to one another under particle-hole).  Indeed such a  topological order was obtained while studying the surface of a particle-hole symmetric  topological insulator (ClassAIII)  \cite{fidkowski3d}. The same anyon content appears in the T-Pfaffian state, a surface topological order proposed for Z$_2$ topological insulators \cite{MetlitskiKaneFisher, fSTO3, fSTO1, Bonderson, mrosscdl15}.  In Ref. \cite{sonphcfl} it was shown that this particle-hole symmetric topological order can be obtained from pairing  Dirac composite fermions\cite{sonphcfl,3dfSPT2,MetlitskiAV,WS16,dualdrcwts2015}  in the s-wave channel, where it was named the PH-Pfaffian (we will adopt this terminology here). On the other hand, pairing in the $d+id$ ($d-id$) channels leads to the Pf (aPf) state, since in the Dirac composite fermion description, time reversal plays the role of particle hole symmetry. Even in the absence of particle hole symmetry the PH-Pfaffian topological order is well defined and Ref. \cite{ssletalapf} had earlier proposed that this same state could arise at the transition between the Pfaffian and anti-Pfaffian phase.

Since the PH-Pfaffian FQH state at  $\nu = 5/2$ is expected to show $K=5/2$,   the experiments may be
considered to be evidence in favor of such a state. However, it has been widely believed, based on numerical calculations, that a state with the quantum numbers of the PH-Pfaffian would have a higher energy than the the Pf or aPf state, and might not even possess an energy gap, for realistic interaction parameters. This belief is partly based on unpublished results and on calculations that focus on the accuracy of the Pf or aPf ground states in spherical geometries but do not explicitly consider the flux numbers appropriate for the PH-Pfaffian state\cite{MorfED, Feiguinetal, StorniMorfDasSarma, RezayiPRL, MorfPrivate}, and it probably deserves further examination. Nevertheless, in order to investigate the effects of disorder, we assume here that the belief is correct.  It was also argued, analytically, in Ref.~\cite{Milovanovic} that a PH-Pfaffian state, when projected into a single Landau level, must necessarily be gapless, but we have not been able to completely follow the logic behind those arguments.

Motivated by previous experiments that seemed to deviate from results expected for the Pf or aPf states,  Zucker and Feldman \cite{feldman} have suggested that the  PH-Pfaffian might somehow be stabilized by disorder, but they were not very specific about how this  might come about. In the current manuscript, we attempt to explore  further the idea that the experimental situation may be a consequence of disorder. We consider a set of related models, with differing assumptions about the energy scales separating the different possible phases and their domain boundaries, as well as the length scale and magnitude of asymmetry fluctuations due to disorder.  We conclude that a disorder-stabilized phase with the quantum numbers of the PH Pfaffian is indeed possible in principle, but the conditions for its realization may be very restrictive.  We believe that it remains  very much an open question whether disorder-induced stabilization is the explanation for the experimentally-observed heat conductivity.

 \begin{center}
\begin{figure}

\captionsetup{justification=raggedright}

\adjustbox{trim={0\width} {0\height} {0\width} {0\height},clip}
{\includegraphics[width=1\columnwidth]{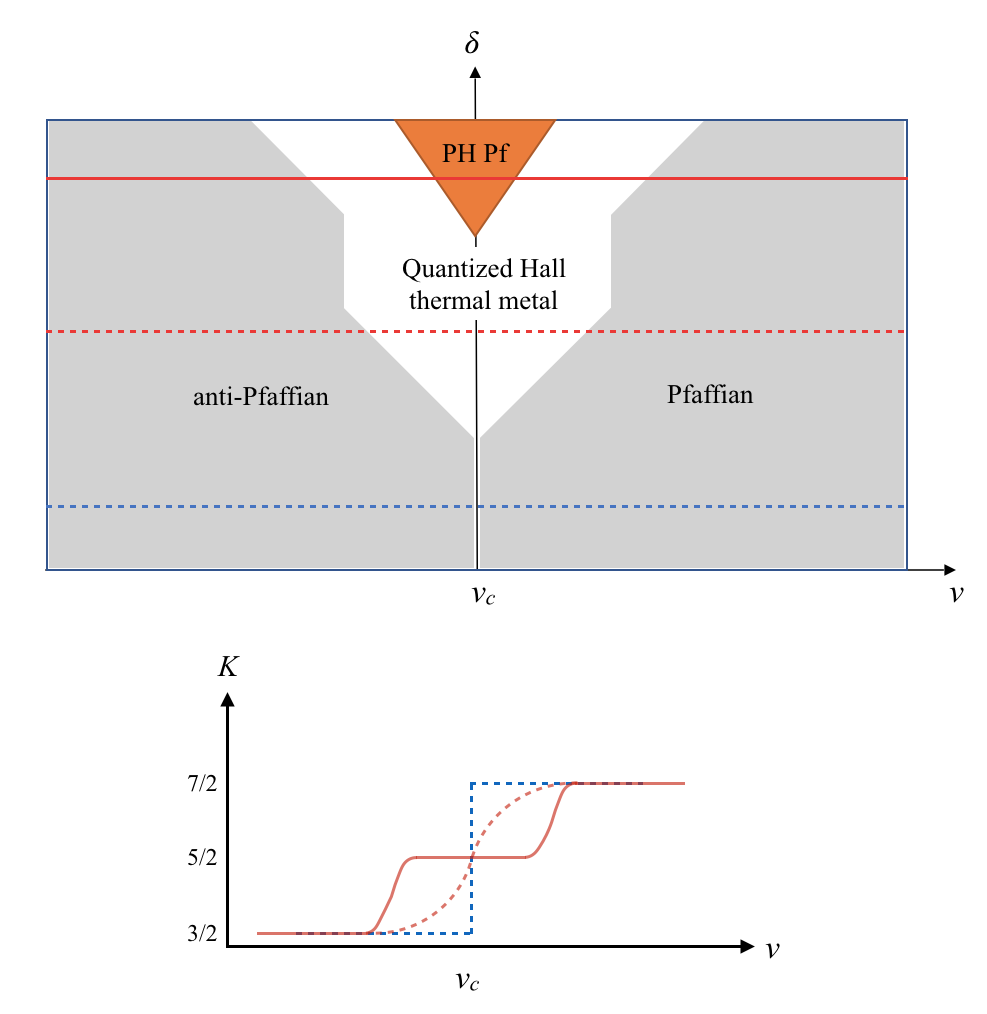}}

\caption{Upper panel: a possible phase diagram containing four phases: Pf, aPf, PH-Pf and the quantized hall thermal metal. The vertical axis $\delta$ is primarily related to disorder strength, but could also be affected by detailed energetics. At a given disorder strength, three scenarios are possible as one varies the filling fraction $\nu$ across the symmetric point $\nu_c$ (which is expected to be shifted slightly away from $5/2$ due to Landau level mixing): a direct first-order-like transition from Pf to aPf phase (dashed blue line), an intermediate thermal metal (dashed red line) and an intermediate PH-Pf phase separated from Pf/aPf by a thermal metal (solid red line). The values of $K$ as functions of $\nu$ in these scenarios are shown in the lower panel.}

\label{PD0}
\end{figure}
\end{center}

Our analysis begins with an assumption that PH asymmetry due to Landau level mixing may be neglected and that the ground state at $\nu=5/2$ in an infinite  disorder-free system may be indifferently  the Pf or aPf state.  We assume that the actual system is subject to potential fluctuations on a length scale $\xi$ large compared to the magnetic length, arising from fluctuations in the density of charged impurities in a doping layer, which is set back from the electron layer. (Typically, $\xi$ will be of the order of the set-back distance.)   We assume that these fluctuations are large enough to favor the nucleation of one or more charged quasiparticles of one or the other sign in a typical region of size $\xi$.  Then, provided that one could neglect the energy cost of a domain wall and one could neglect the thickness of such a wall, 
one might expect that  sample would be divided into Pf and aPf regions, with a characteristic size scale of order $\xi$, and roughly equal amounts of each phase.  In this scenario, as we shall discuss further below,  if one can neglect tunneling between domain walls that come close together, the macroscopic thermal Hall conductance  would be either $K = (7/2) $ or $K = (3/2) $ depending on which of the two phases percolates across the sample.  Consequently, in this model, if one were to vary the average filling fraction by varying magnetic field, one would expect a sharp transition between the two values of $K$ as first one, then the other phase percolates. This is illustrated by the dashed blue curve in Fig.~\ref{PD0}.

\begin{figure}
 \begin{center}
\captionsetup{justification=raggedright}

\adjustbox{trim={0.12\width} {0\height} {0\width} {0\height},clip}
{\includegraphics[width=1.3\columnwidth]{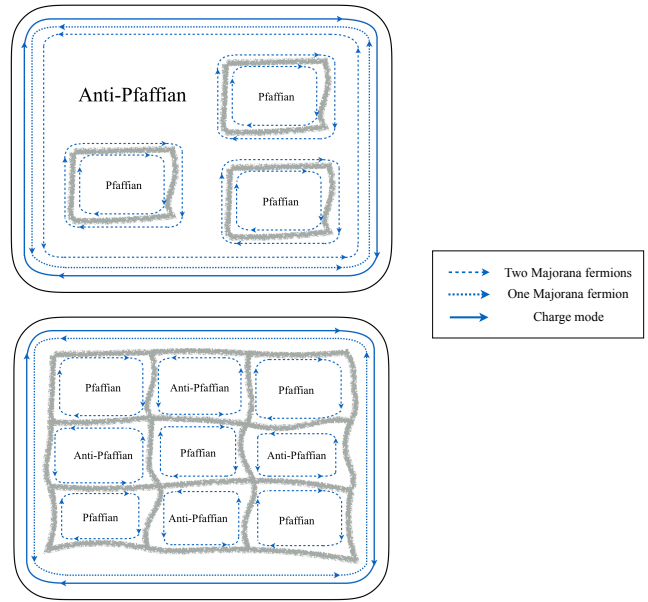}}

\caption{Illustrations of inhomogeneous Pfaffian/anti-Pfaffian mixtures. The upper panel describes when a single phase (for example anti-Pfaffian) percolates, and the system behaves on macroscopic scales as the percolated phase, with the corresponding edge states. The lower panel describes when neither of the two phases percolates, but the domain walls between them percolate. If the neutral modes propagate diffusively on the domain walls, the system becomes a thermal metal (but is still a quantized electric Hall insulator). If the neutral modes are localized instead, the system becomes equivalent to a PH-Pfaffian state on macroscopic scales, with the edge modes that give $K=5/2$.
}
\label{evolve}
\end{center}
\end{figure}

Other possibilities arise, however,  if we assume that the domain walls have a finite thickness, but we still neglect the domain wall energy cost.  By domain wall thickness, we mean that if two domain walls become closer together than the domain wall thickness,  we must take into account the  possibility of the tunneling of excitations between the two domain walls. Then we have the possibility that for a finite range of  $\nu$, neither Pf nor aPf will percolate; rather the domain walls will form the percolating phase (see Fig.~\ref{evolve}). Then  we should consider at least two possible outcomes.  One is that low energy neutral excitations can diffuse freely on the network of domain walls, the second is that only localized excitations can exist at low energies.   We argue below  that in the latter case, the system will have a quantized thermal Hall conductance with $K = (5/2) $, and will share the topological properties of the PH-Pfaffian phase. The state with delocalized neutral excitations, which we denote as the ``quantized Hall thermal metal" (QHTM) state,   will not have a quantized thermal Hall conductance, but may have a value of $K$ that varies continuously as one varies $\nu$. The metallic phase will also have a longitudinal thermal conductance  $\kappa_{xx}$ which differs from zero and which would have to be taken into account in analyzing the experiments, which actually measure a two-terminal thermal conductance.  However, we emphasize that the QHTM is still an (electrically insulating) quantized Hall state, whose electrical Hall conductivity $\sigma_{xy}=(5/2)  e^2/h$  remains quantized, and $\sigma_{xx}=0$ at zero temperature, since charged quasiparticles should still be localized by the disorder and Coulomb interactions. The two scenarios are both  illustrated, in Fig.~\ref{PD0},  by the red solid and dotted curves. 

Even without disorder, a highly anisotropic version of the quantized Hall thermal metal at $\nu=5/2$ has been proposed earlier in Ref.~\cite{WanYang} as a candidate phase due to spontaneous formation of Pf and aPf stripes. Here we focus on the scenario in which the system is a pure Pf or aPf state in the clean limit, and the quantized Hall thermal metal phase, if realized, is a result of disorder.

We note that the localization problem studied here will fall into the Altlander-Zirnbauer symmetry class D\cite{AZ}, involving fermions without any nontrivial global symmetry. In general, systems in this class can be either localized or delocalized \cite{SenthilFisherD, DMetal}, depending on microscopic details.  Analysis of these details will be a principal focus of the present paper.  

It is important to note that in practice, the geometric domain structure will be driven by energy considerations, and will certainly not be completely random.   For example we must take into account the energy cost of domain walls.  If this energy cost is too high, or if the domain size $\xi$ is too large relative to the domain wall thickness, it may be energetically favorable  for the system to spontaneously minimize the length of domain walls by arranging itself so that either the Pf  phase or the aPf phase percolates, rather than the domain walls,  at any  value of the magnetic field.  In this case, we return to the scenario of the blue curve in Fig~\ref{PD0}, which is to say a direct first-order transition between the Pf and aPf states.  

It has been argued that in two dimensions, a first-order transition should, in principle, be broadened in the presence of disorder, by an amount of order $exp (-1 / \delta^2)$, where $\delta$ is proportional to the root-mean-square strength of the fluctuating disorder potential\cite{ImryMa,Binder1983}.  It is not clear whether these arguments apply to the present situation, but in any case, the broadening  of the transition should be negligibly small in the case of weak disorder. In a finite sample, the precise transition point may depend on boundary conditions, and will generally vary slightly from one realization to another. Also, there can be a small range of parameters where both phases can percolate across the sample in one direction. We neglect such finite-size effects in our discussions.  

If the disorder potential is weak, then only rare fluctuations will be large enough to produce localized quasiparticles of either sign.  In this case, the length scale for possible formation of Pf and aPf domains should become large. Then energy cost of domain walls should be dominant, and the first-order-like transition scenario described above should be favored.

We present  a schematic phase diagram in Fig.~\ref{PD0}, which describes a possible  scenario  consistent with the above discussion. If this phase diagram is correct, and if the phase observed in experiments is indeed a PH-Pfaffian stabilized by disorder, then there is an interesting prediction: with decreasing disorder, the system would first enter the QHTM phase, with unquantized thermal conductance, before landing on either the Pf or aPf phase in the limit of very weak disorder.

In all of  the above scenarios, as previously mentioned, the electrical Hall conductivity will remain quantized at the value $(5/2)  e^2/h$, and the longitudinal conductance $\sigma_{xx}$  will remain zero, in the limit of zero temperature, as long as the charged quasiparticles remain localized.   
Of course, if the filling fraction deviates too far from $5/2$, so that the density of quasiparticles becomes too large, the quantized conductance will be lost, either because the  quasiparticles are delocalized and are free to move, or because there is a transition to a completely different phase. 
If the region of quantized electrical conductance is too small, then one would not be able access more than one of the phases shown in Fig.~\ref{PD0}, and one might then observe only one value of the thermal Hall conductance in the region where the electrical Hall conductance is pinned on the 5/2 plateau.

 %In addition to a possible mechanism for disorder generated PH-Pfaffian state, we find that a natural phase that appears in this setting is  the QHTM, as shown in the figure. In particular, if we assume that the $K=5/2$ state observed experimentally is indeed a disorder-stabilized PH-Pfaffian state, then we are lead to a very natural prediction: if one keeps reducing the level of disorder at a fixed filling fraction $\nu$ (inside the electric Hall plateau), at some point the system will enter the QHTM phase, with un-quantized $\kappa_{xy}/T$ and nonzero $\kappa_{xx}/T$. In the limit of very weak disorder the system will eventually land on either the Pfaffian or anti-Pfaffian phase. 

 In the following three sections, we shall formulate  and analyze a network model designed to study the possible phases that can result if a  first-order-like transition can be avoided, and a percolating network of domain walls can be achieved.  In the two subsequent sections, we include some additional remarks, and we restate our conclusions.

 \section { Structure of the domain walls and junctions}

Our goal here is to analyze the situation  where  domain walls are allowed to intersect, potentially leading to either the second or third scenario defined above.  We assume that  in the interior of any Pf or aPf   region there is a non-zero energy gap for mobile excitations, so that heat conduction at low temperatures is determined by the low-energy modes that propagate along  the domain boundaries.  Our first task will be to characterize these propagating modes.  Then, we shall discuss the possible ways that these modes can be connected at a junction between two domain walls.

 \subsection {Domain wall structure}
 \label{domainwallsubsec}
 
 From topological considerations, we know that the boundary between a Pf and aPf domain must carry a thermal conductance with $K = 2 $.   Quite generally, this may be described in terms of four co-propagating  chiral Majorana modes, each of which has $K =  1/2$. However, at least in the simplest cases, we find that the four Majorana modes occur in pairs that may be combined to give two complex chiral fermion modes, generically with different velocities.  
 
 To model a domain boundary, we assume that the Pf and APF phases can be represented as paired superconducting states formed from a Fermi sea of composite fermions.
  We employ specifically  the formalism introduced by D. T. Son \cite{sonphcfl}, in which the system is described by a set of massless Dirac fermions, interacting with an emergent gauge field $a_\mu$.  The Hamiltonian for the fermions may be written as  $H=H_0 + H_\Delta$, where
 
 \be
 H_0 =   \int d^2r \, \psi^\dagger  (v_F  \vec {k}  \cdot \vec{\sigma} - \mu) \psi , 
 \ee
 \be
 \label{HDelta} 
 H_\Delta = \int d^2 r  \sum_l  k_F^{-| l|}  \psi_\uparrow  \{ \Delta_l ^*,   (k_x + i k_y) ^l  \}   \psi_\downarrow  + \rm{h. c.}
 \ee
 Here $\vec{k} = \vec{p}-\vec{a}$, where $\vec{p} = - i \hbar \nabla$,  the chemical potential $\mu$ is a constant, and $\{\, , \, \}$ is the anitcommutator.  We shall consider a uniform domain wall which is oriented along  the x--axis and centered at $y=0$, so that the pairing potentials $\Delta_l $ are functions of $y$ but are independent of $x$ Moreover, we consider a situation in which $\vec{a}=0$.  
 
 In principle, the pairing potentials should be determined self-consistently, by equations which have the form, in a local approximation, of 
 \be
 \Delta _l  \propto  - \lambda_l \,  \langle \psi_\uparrow    (k_x + i k_y) ^l     \psi_\downarrow \rangle .  
 \ee
 We shall assume here that $\lambda_l=0$ for all $l$ other than $l=\pm 2$, and that $\lambda_2 = \lambda_{-2} > 0$.  Thus, only $l= \pm 2$ will enter in Eq. (\ref{HDelta}). 
 We shall not actually solve any self-consistency equations here, but shall only make use of the qualitative form of $\Delta_l$. We shall assume that $|\Delta|$ is small compared to the Fermi energy, $E_F = v_F k_F$, but not necessarily very small. 
 
 To describe a domain wall, we assume  that $\Delta_2 \to 0$, for $y \to - \infty$, and $\Delta_{-2} \to 0$, for $y \to + \infty$.  Further, we assume that $\Delta_2$ and $\Delta_{-2}$    go to  constants for  $y \to \pm \infty$, respectively, with equal magnitudes  given by the bulk order parameter, and with specified phases, $\phi_2$ and $\phi_{-2}$.   Placing the center of the domain wall at $y=0$, we may assume by symmetry that 
 $|\Delta_2(y)| = |\Delta_{-2} (-y)|$, so the two order parameters have equal values at $y=0$.  If we neglect the spatial variation of the order parameters about $y=0$, we then find that the energy spectrum there has four Weyl nodes, located  at points with $k_x + i k_y = k_F e^{i \theta_j}$, with 
 \be
 4 \theta_j = \pi - \delta \phi + 2 \pi j ,
 \ee 
where $j$ goes from 1 to 4, and  $\delta \phi \equiv \phi_2 - \phi_{-2}$. 
 
If the spatial variation of the order parameter is slow compared to $1/k_F$, we can neglect mixing between the nodes, and the effect is that of a mass parameter that varies linearly with $y$ in the vicinity of $y=0$. This model can be solved by standard methods\cite{JackiwRebbi}, and one finds that each of the four nodes gives rise to a one-dimensional  chiral Majorana mode, with a linear dispersion along the direction of the domain wall, in addition to some number of massive fermion modes, which we  neglect here, assuming that  the temperature is sufficiently low. This domain wall structure with four chiral Majorana fermions was discussed in Ref.~\cite{BarkeshliMulliganFisher}. The velocities  of the  Majorana modes are given by 
\be
\label{vj} 
v_j = ( v_F^2 \cos^2  \theta_j + v_{\Delta}^2 \sin^2 \theta_j) ^{1/2} ,
\ee  
where 
\be
v_{\Delta} = 4 |\Delta_2| / k_F ,
\ee
and $\Delta_2$, here,  is the value of the $l=2$ order parameter at $y=0$. For a domain wall oriented at an angle $\alpha$ relative to the x-axis, the arguments  $\theta_j$ in Eq. (\ref{vj}) should be replaced by $(\theta_j - \alpha)$. 

Because the four nodes $\theta_j$ are separated in momentum by an amount of order $k_F$,  disorder with wave vectors small compared to $k_F$ will not be effective in mixing the Majorana modes, which originate from different nodes.  Also, if the domain wall  changes its direction on a length scale large compared to $1/k_F$, the Majorana modes should follow along adiabatically, without significant mixing.  

Perturbations that can cause scattering between the Majorana modes can occur, however, if there is a localized charged quasiparticle very close to the domain wall.  Quasiparticles are associated with vortices in the superconducting order parameter, which can cause large momentum transfers for composite fermions close to the vortex core.    

We remark that the phase difference $\delta \phi$ is expected to be constant, or at most a slowly varying function of position, along any connected network of domain walls   This is because the order parameters $\Delta_2$ and $\Delta_{-2}$ couple to the same vector potential $\vec{a}$, and to minimize the energy we require that $(\nabla \phi_2 -\vec{a)}$ and
$(\nabla \phi_{-2} -\vec{a})$  should both vanish in any region where both order parameters are different from zero, so the  phase gradients must be equal.

\subsection {Junction between domain walls}  

\begin{figure}
\begin{center}

%\captionsetup{justification=raggedright}

%\adjustbox{trim={0\width} {0\height} {0\width} {0\height},clip}

\begin{subfigure}{0.48\columnwidth}
\includegraphics[width=0.7\columnwidth]{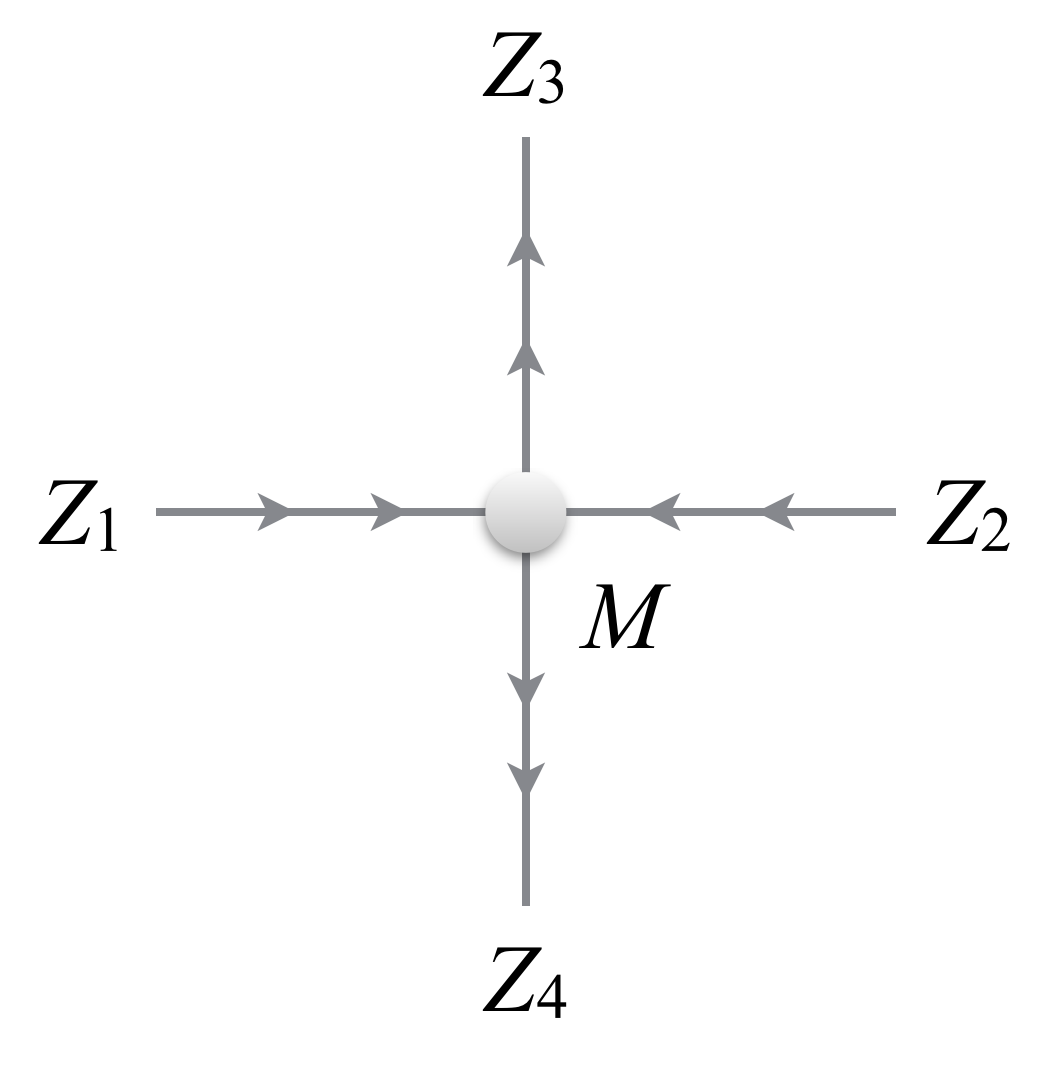}
\caption{}
\label{junction0}
\end{subfigure} \hspace{.5pt}
\begin{subfigure}{0.48\columnwidth}
\includegraphics[width=0.7\columnwidth]{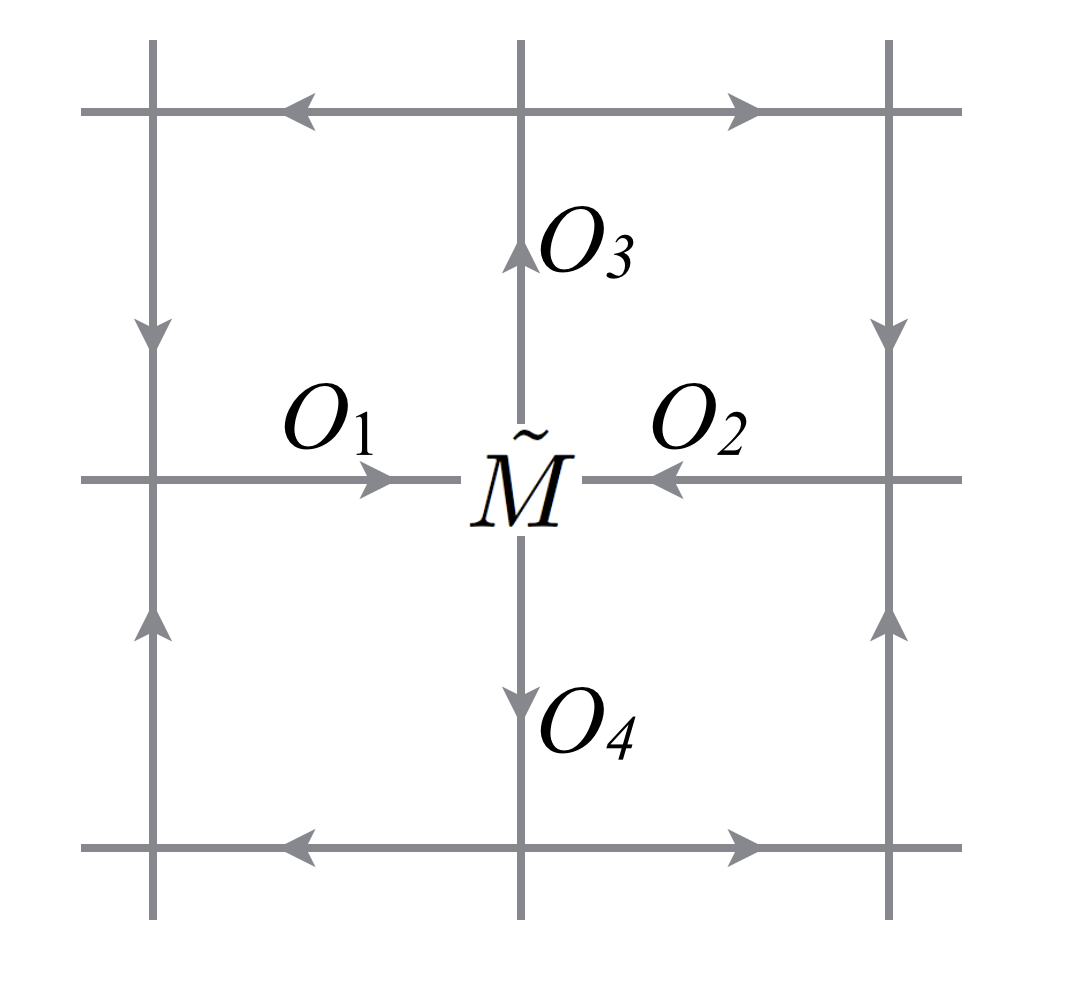}
\caption{}
\label{network}
\end{subfigure}

%\caption{}

\caption{(a) Scattering at each domain wall junction, described by the scattering matrix $M$. (b) Illustration of the network model. Associated with each link is an $O(4)$ flavor mixing matrix, and with each vertex a scattering matrix $\tilde{M}$.}
%\label{junction0}
\end{center}
\end{figure}

We now consider what happens if two domain walls cross at a point, as indicated in Figure~\ref{junction0}. The junction is here oriented so that there are four Majorana modes each entering from the left and from the right, and four modes, each, leaving   towards the top and towards the bottom.  Transmission through the junction can be described by a scattering matrix of the form 

\be
\lp \begin{array}{c} Z_3 \\ Z_4 \end{array}\rp=M\lp \begin{array}{c} Z_1 \\ Z_2 \end{array}\rp,
\ee
where $Z_1$ and $Z_2$ ($Z_3$ and $Z_4$) are the four-component real amplitudes of the incoming (outgoing) Majorana fermions.

%\begin{center}
%\end{center}

If there are no charge quasiparticles close to the junction, we may assume that the two order parameters vary smoothly near the junction, and that the difference $|\Delta_2|^2 - |\Delta_{-2}|^2$ vanishes quadratically near the center.   (For the geometry illustrated,  this difference is proportional  to $xy$.)   Then there should, again,  be no mixing between the four Majorana  ``flavors", corresponding to the four nodes $\theta_j$, and $M$ will be block-diagonal in the flavor space.  More generally, if we take into account the possibility of scattering between the four modes on any given domain wall, we may write $M$ in the general form 
\be
M=O_1O_3(\oplus_jM_j)O_4O_2,
\ee
where $O_a$ is an $O(4)$ rotation matrix acting on $Z_a$ ($a=1,2,3,4$) that describes the mixing of the four flavors of Majorana fermions on link $a$, and the reduced scattering matrix $(\oplus_iM_j)\equiv \tilde{M}$ is block-diagonal in the flavor space ($j$ being the flavor index). Unitarity requires the real matrix $M_i$ to take the form
\be
M_j=\lp \begin{array}{cc} \cos\alpha_j & \sin\alpha_j \\ -\sin\alpha_j & \cos\alpha_j \end{array}\rp.
\ee
The probability amplitude for mode $i$ to turn left and right is given by $\cos \alpha_i$ and $\sin \alpha_i$, respectively. The critical point is at $\alpha_c=\pi/4$.

Formally we can absorb the flavor-mixing matrices $O_a$ to the links and think of $\tilde{M}$ as the scattering matrices at the vertices, even if physically the flavor-mixing can happen at both the links  and the vertices.

 \section {Chalker-Coddington-type analysis}
 \label{networksection}

\subsection{Model setup}
\label{model}

Here we analyze the localization problem of the Majorana fermions on the Pfaffian/anti-Pfaffian domain walls, assuming that the domain walls (instead of the Pfaffian or anti-Pfaffian domains) percolate over the entire sample. The analysis is carried out using a Chalker-Coddington type of network model, where one assumes a regular geometry for the domain walls,  as illustrated in Fig.~\ref{network}, but introduces randomness for the propagators along the bonds and/or for the unitary matrices at the junctions. \cite{ChalkerCoddington}
 A closely related model with only one Majorana fermions per link has been studied previously in Ref.~\cite{DMetal}.  At each vertex of the network there is a scattering matrix $\tilde{M}$ as defined in previous section, and along each link we introduce a flavor mixing matrix $O_a$, which takes into account any flavor mixing due to scattering along the link, as well as rotations accumulate during passage along the link due to any differences in the wave vectors associated with the different modes.  In the simulations described below, we 
assume that $\tilde{M}$ is spatially uniform, and put all the randomness in $O_a$.

Numerically we simulate the model on a quasi one-dimensional $L\times N$ cylinder, where $N$ can be large (maximally we go to $5\times 10^4$) and $L$ is relatively small (maximally we go to $32$). The total transfer matrix describing propagations across the entire cylinder (in the $N$ direction) is obtained by multiplying up all the scattering matrices on the vertices and the mixing matrices on the links. The localization length $\xi_L$ of this quasi-$1D$ system is then extracted using standard methods\cite{ChalkerCoddington, MacKinnon1983}. Whether the two dimensional system is localized or not is determined by whether $\xi_L$ saturates as $L$ grows, or equivalently, whether $\Lambda_L\equiv\xi_L/L$ decreases as $L$ grows.

There are two groups of parameters that determine the model: the uniform part, given by $\alpha_i$, and the random part, given by the form of the random matrices $O_a$ (namely how random they are allowed to be).

For each fermion flavor, at $\alpha_c=\pi/4$ the vertex is symmetric between left-turning and right-turning. In the clean limit of the network model ($O_a=1$) this simply gives a gapless Majorana fermion describing the transition between a trivial and a $p+ip$ topological superconductor. Away from this special value of $\alpha_c=\pi/4$  the Majorana fermion becomes gapped, and time-reversal symmetry (now broken) relates $\alpha$ to $\alpha'=\pi/2-\alpha$, corresponding to positive and negative Majorana mass, respectively (again in the clean limit of the network model).

We now discuss the form of randomness on $O_a$. There are roughly three classes of randomness here:
\begin{enumerate}
\item The fermions can mix due to random scattering, which comes from a unitary process and gives an $O_a\in SO(4)$ (not $O(4)$). As discussed in Sec.~\ref{domainwallsubsec} this requires short-range disorder, which is in principle different from the longer-ranged disorder that is primarily responsible for the formation of the domain walls. We implement this on each link as $e^{\phi_{i,j}T_{i,j}}$, where $T_{i,j}$ is one of the six generators of $SO(4)$ that rotates in the $(i,j)$ plane ($1\leq i< j\leq 4$), and $i,j$ are picked randomly at each link, giving rise to mixing of two flavors chosen randomly. $\phi_a$ is a random number uniformly distributed in $(-\eta_{i,j}\pi,\eta_{i,j}\pi)$ with some number $\eta_{i,j}$ controlling the strength of scattering in the channel given by the generator $T_{i,j}$.  Strong scattering corresponds to $\eta=1$  where the two flavors can mix arbitrarily, while a small $\eta$ means weak scattering (mixing). One can implement an even stronger mixing by first generating several independent $O\in SO(4)$ and then taking their product.  We expect that this will not change the results qualitatively, though we will use a version of this in Sec.~\ref{ScB} for convenience.

\item Random vortices ($\pi$-fluxes) within the domains. In our physical context they correspond to charge $\pm e/4$ quasi-particles nucleated by impurity potentials, and therefore should be included in general unless there exist some highly nontrivial mechanism to suppress unpaired vortex excitations within the domains. Notice that a real $e/4$ quasi-particle (vortex) comes with a localized Majorana zero mode associated with it, and in principle this zero mode can mix with the fermions on the domain walls. But since the domain size should be much larger than the correlation length (otherwise it could hardly be called a domain), the vortices will typically be deep within the domains and therefore their  zero modes will not mix with the domain wall fermions. Therefore we do not attempt to include such zero modes in our network model. In our model an odd number of vortices within a domain correspond to a nontrivial flux $O_1O_2O_3O_4=-1$ on a plaquette ($1,2,3,4$ being the edges of the plaquette), and therefore random vorticity is implemented through random signs on each link that are seen by all the four flavors, i.e. $O_a=\pm 1$. Physically it seems natural to assume that half of the domains contain odd number of vortices, so the random signs come with equal probability. One can also consider more general cases with different average vortex density, implemented in the model by having probability $p$ for $O_a=-1$ on each link. Notice that this type of randomness is already contained in the $SO(4)$ mixing from random scattering, unless there is at least one flavor $i$ that is only weakly mixing with any other flavor ($\eta_{i,j}\ll1$ for some $i$ and any $j$), in which case this random vorticity disorder has to be included separately. 
 
\item Random ``improper" vortices: these are $\pi$-vortices that are seen only by one or three out of the four Majorana fermions. Formally this enlarges the space of $O_a$ from $SO(4)$ to $O(4)$. Physically they corresponds to vortices that lie in between the Majorana fermions (assuming they have some spatial separation). However such vortices should be (1) very rare, and (2) energetically unstable toward redrawing the domain walls. We therefore assume that they are irrelevant for realistic systems. One can still study them as a theoretical problem, and we find that they almost always lead to a thermal metal (delocalized) phase, in agreement with earlier works with only one flavor of Majorana fermion\cite{DMetal}. We do not include improper vortices in any calculations described below.
\end{enumerate}

%We will call the above disorders type-1, type-2, and type-3 in the subsequent discussions.

\section {Results of the network model}

As we shall see below, our network model does not behave in a universal manner (in contrast to the usual Chalker-Coddington model describing quantum Hall plateau transitions). Instead the long-distance behaviors depend heavily on detailed choices of various parameters, which in turn depend on microscopic details including energetics of the domain walls and junctions. Given that we understand very little about the exact structures of the domain walls and junctions, we will discuss several distinct scenarios below, in decreasing order of ``naturalness" -- a scenario is considered more natural if it involves less nontrivial assumptions about the domain walls. We find that in the scenario with minimal assumption, there is only a metallic phase between the Pfaffian and anti-Pfaffian states at finite disorder, which is at odds with the experimental observation of $K=5/2$. A PH-Pfaffian state could emerge (with $K=5/2$) in some other scenarios if we make certain extra assumptions about the domain wall structures. Among the simpler ones of those scenarios, the PH-Pfaffian state and the Pfaffian/anti-Pfaffian states are separated by an intervening thermal metal. In such cases we predict that if one further reduce the disorder strength, the system would go through a thermal metal phase with nonzero $\kappa_{xx}/T$, before eventually going into either the Pfaffian or anti-Pfaffian state in the clean limit.

\subsection{Scenario A: an intervening metal (and nothing else)}

We first look at the simplest scenario, which as we shall see does not explain experimental measurements. In this scenario we do not assume any special structures between the four Majorana fermions on the domain walls. In particular, we assume that the scattering (or tunneling) at the domain wall junctions (vertices in the network model) does not discriminate different flavors. For the network model this means that $\eta_{i,j}=\eta$ is identical for all $(i,j)$, and $\alpha_i=\alpha$ is identical for the four flavors. As one tunes the system from the Pfaffian to the anti-Pfaffian state, the value of $\alpha$ turns from $\alpha\ll\alpha_c$ to $\alpha\gg\alpha_c$. In the clean limit of the network model ($O_a=1$) this simply describes four Majorana fermions with identical mass gaps.

\begin{center}
\begin{figure}

\captionsetup{justification=raggedright}

\adjustbox{trim={0\width} {0\height} {0\width} {0\height},clip}
{\includegraphics[width=0.8\columnwidth]{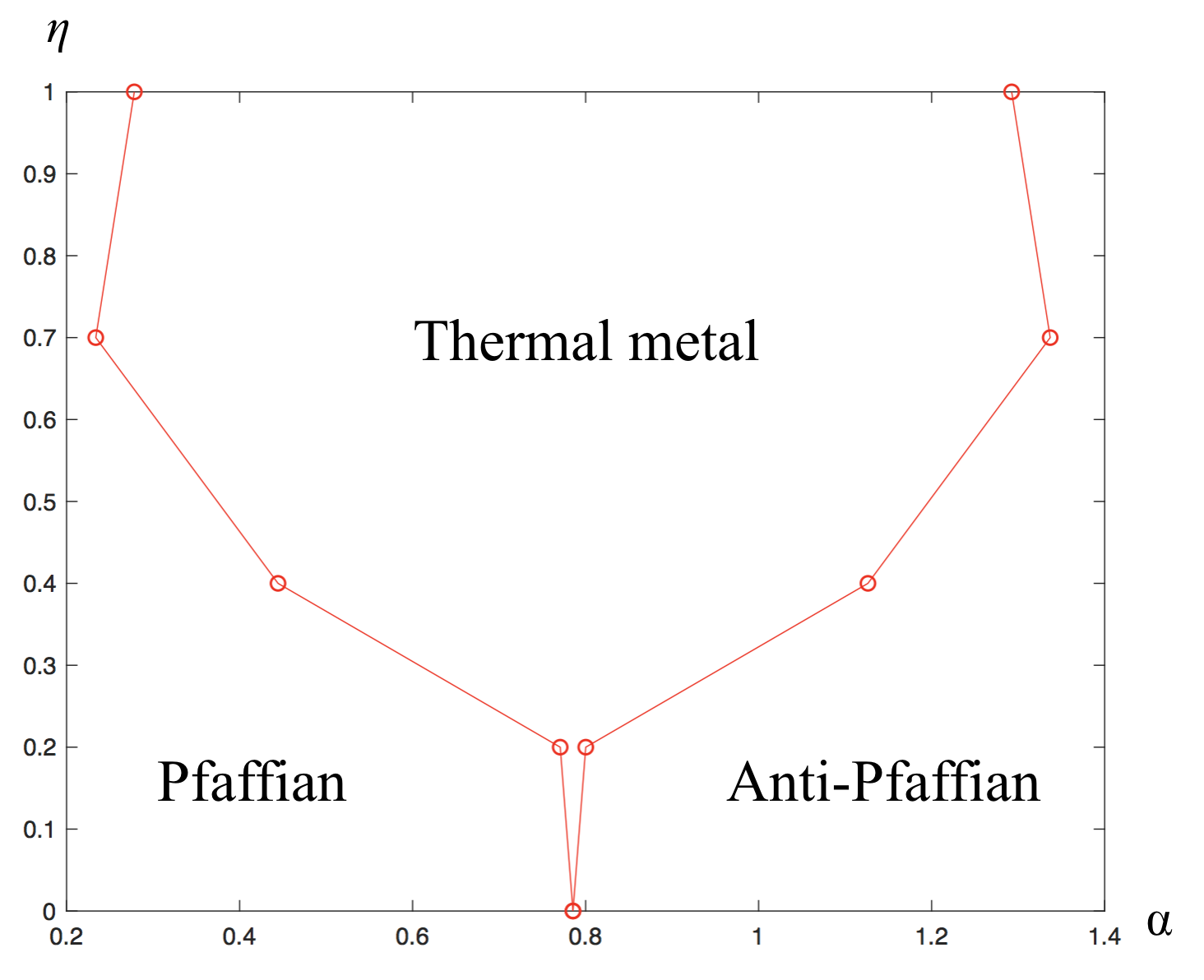}}

\caption{Phase diagram of the network model in Scenario A, with $\eta_{i,j}=\eta$ for any $(i,j)$ and $\alpha_i=\alpha$ for any $i$. The Pfaffian/anti-Pfaffian transition in the clean limit ($\eta\to0$ and $\alpha=\pi/4$) is broadened to a thermal metal phase upon introducing a generic disorder. Here, as in Fig.~\ref{LambdaPH} and \ref{PD2}, points represent data, while lines are guides for the eye.
}
\label{PD1}
\end{figure}
\end{center}

Numerically we find that, in this scenario the Pfaffian/anti-Pfaffian states are stable as long as $\alpha$ is sufficiently far away from $\alpha_c$. When $\alpha$ becomes close (not necessarily identical) to $\alpha_c$, the system becomes delocalized and we obtain a thermal metal. The metallic region broadens as $\eta$ grows. There is no other localized phase in between. A crude phase diagram is shown in Fig.~\ref{PD1}. Notice that for the real physical system, $\alpha$ is related but not directly identified with $\nu$: if disorder is too weak, the transition between Pf and aPf states is first-order-like, and we expect a discontinuous jump in $\alpha$ as one tunes $\nu$ across $\nu_c$. Only in the intermediate disorder regime we expect $(\alpha-\alpha_c)\sim(\nu-\nu_c) $.

Random vortices, which should be included for real systems, were omitted in the above consideration, except in the limit with strong $\eta$, where random vortices are automatically included. Adding random vortices in our model will significantly broaden the metallic region at small $\eta$. In fact when $\eta\to0$, it was already noted in Ref.~\cite{DMetal} that a finite density of uncorrelated random vortices makes the system metallic for any $\alpha$ (except at the two extreme points $\alpha=0$ or $\pi/2$). This, however, does not change our conclusion for the real system at clean limit, which is described by a first-order-like switching between Pf and aPf phases.
 
Notice that the broadening of the transition into a metallic phase is in sharp contrast with some other symmetry classes of disordered free fermion systems that do not allow metallic phases. For example, if we have a clean system realizing a direct continuous transition between a $\nu=1$ and $\nu=-1$ integer quantum hall phases (symmetry class A, with only charge conservation), once disorder is introduced, the critical point will be localized to a stable phase ($\nu=0$ insulator), with two transitions nearby, one from $\nu=1$  to $\nu=0$ and the other from $\nu=0$ to $\nu=-1$. However, in our case (class D) since a metal is allowed, the critical point between the Pfaffian and anti-Pfaffian states simply broadens to a metal upon introducing disorder. From the scaling theory point of view, this means that the transition point in the clean limit has $\kappa_{xx}/T$ that is already greater than that at the metal-insulator transition, so introducing disorder will only make it more metallic. We notice that the broadening of similar transitions to thermal metal phases under sufficiently generic disorder was already discussed in Ref.~\cite{readgrn}.

If this scenario holds, there should be no state between Pfaffian and anti-Pfaffian with $K=5/2$ unless the system happens to be fine-tuned, which is in tension with the experimental findings. We therefore switch to other scenarios with additional assumptions on the domain wall structures. 

\subsection{Scenario B: PH-Pfaffian and thermal metal} 
\label{ScB}
 
\begin{center}
\begin{figure}

\captionsetup{justification=raggedright}

\adjustbox{trim={0\width} {0\height} {0\width} {0\height},clip}
{\includegraphics[width=0.6\columnwidth]{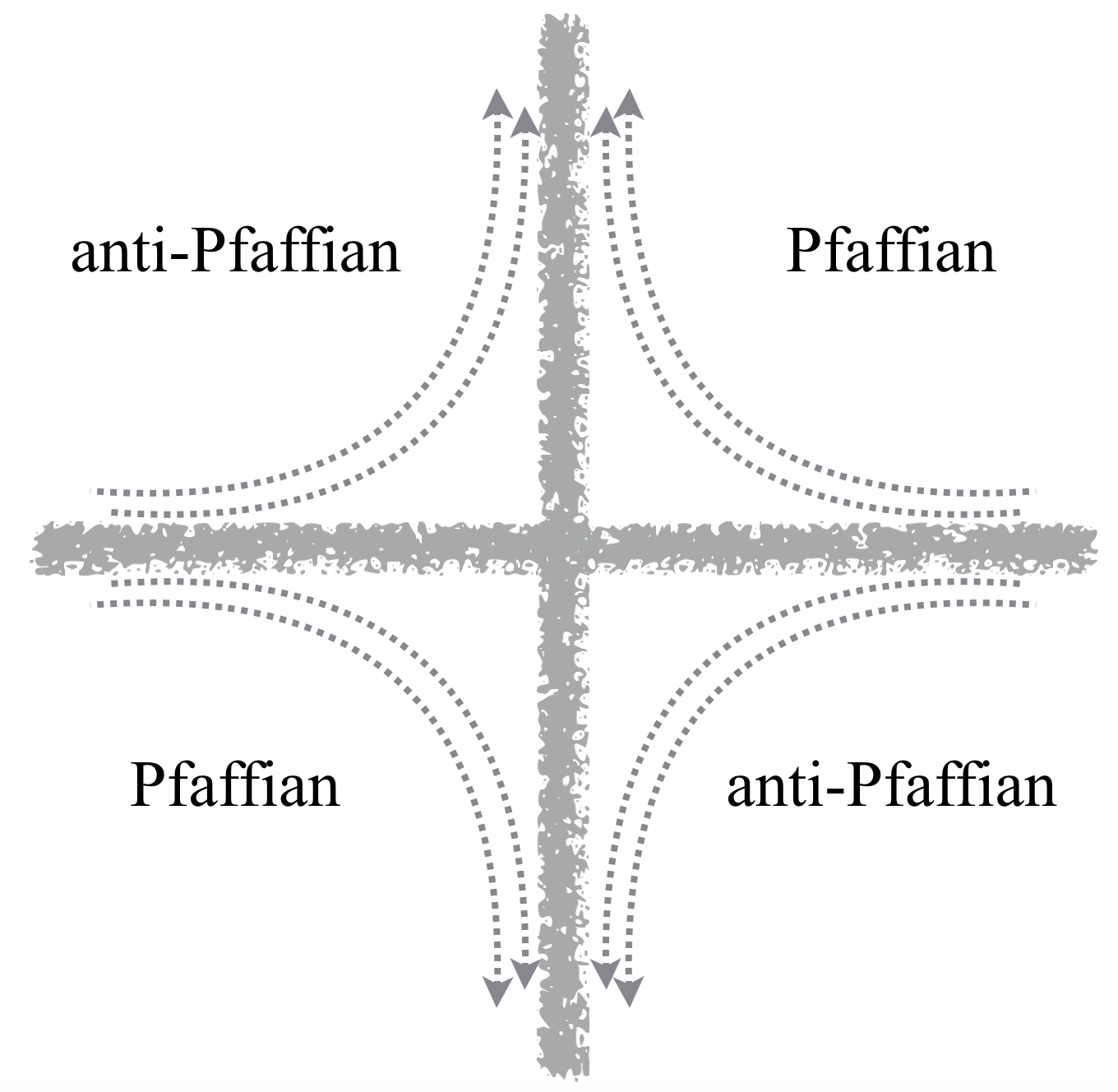}}

\caption{A PH-Pfaffian state can be stabilized if we assume the the two pairs of Majorana chiral modes on each domain wall are spatially separated.
}
\label{gappedwall}
\end{figure}
\end{center}

We now consider the possibility that the four chiral Majorana modes on a domain wall are spatially separated into two pairs, as shown in Fig.~\ref{gappedwall}. This could happen if, for example, a sliver of $s$-wave pairing (giving rise to PH-Pfaffian state) appears within the domain wall. We do not have a satisfactory justification for why this would happen, except to note that it is not forbidden (or ruled out by existing evidence), and we shall proceed without further justification of this assumption. The scattering at the junction will acquire a nontrivial structure due to the spatial separation: each pair will predominantly be scattered to the ``right pair" of modes through the junction. In the network model this means that $\eta_{i,j}\ll1$ unless $(i,j)$ is $(1,3)$ or $(2,4)$. In fact we can implement a slightly modified model that will turn out to be more convenient: on each link $a$ we have $O_a=O_1O_2O_3$ where $O_1=e^{\phi'_{1,3}T_{1,3}}$, $O_2=e^{\phi'_{2,4}T_{2,4}}$ and $O_3=e^{\phi_{i,j}T_{i,j}}$ where $(i,j)$ are picked randomly among all six possible pairs. We take $\phi_{i,j}\in(-\eta\pi,\eta\pi)$ with $\eta\ll1$ independent of $(i,j)$, while $\phi'_{1,3},\phi'_{2,4}\in (-\eta_{intra}\pi,\eta_{intra}\pi)$ with a different constant $\eta_{intra}$ that can be much larger than $\eta$. The intra-pair scattering ($(1,3)$ and $(2,4)$) are greatly enhanced in this model.

At a fixed finite disorder strength that is not too small, as one tunes $\nu$ starting from the anti-Pfaffian state to a Pfaffian/anti-Pfaffian mixture (Fig.~\ref{evolve}), one of the two pairs of Majorana modes turn from predominantly right-turning to predominantly left-turning. This is modeled in the network model as having $\alpha_1=\alpha_3$, $\alpha_2=\alpha_4$, and we tune $\alpha_{average}=(\sum_i\alpha_i)/4$ while keeping $\Delta\alpha=\alpha_2-\alpha_1$ fixed.

\begin{center}
\begin{figure}

\captionsetup{justification=raggedright}

\adjustbox{trim={0\width} {0\height} {0\width} {0\height},clip}
{\includegraphics[width=.9\columnwidth]{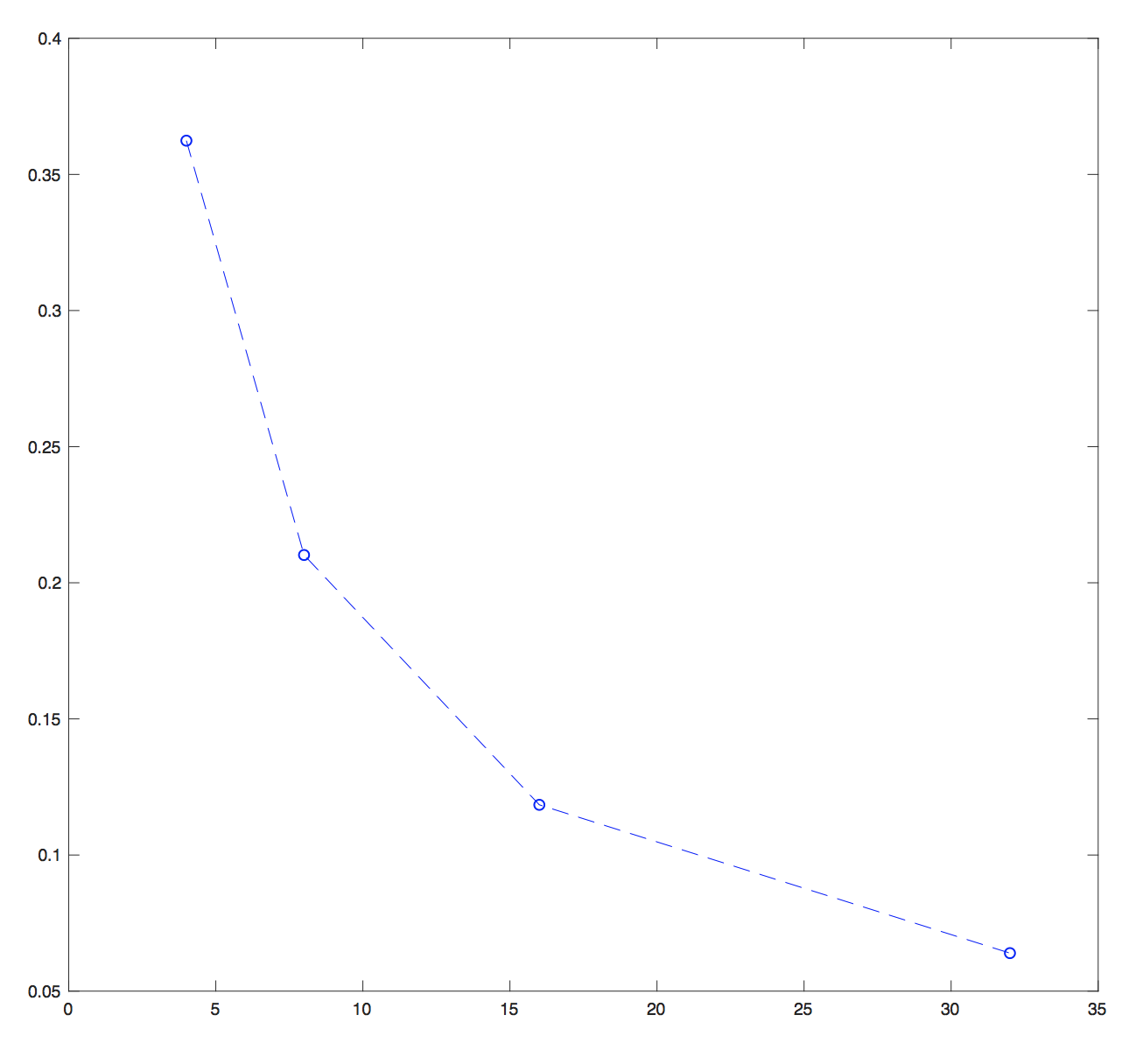}}
{\includegraphics[width=.8\columnwidth]{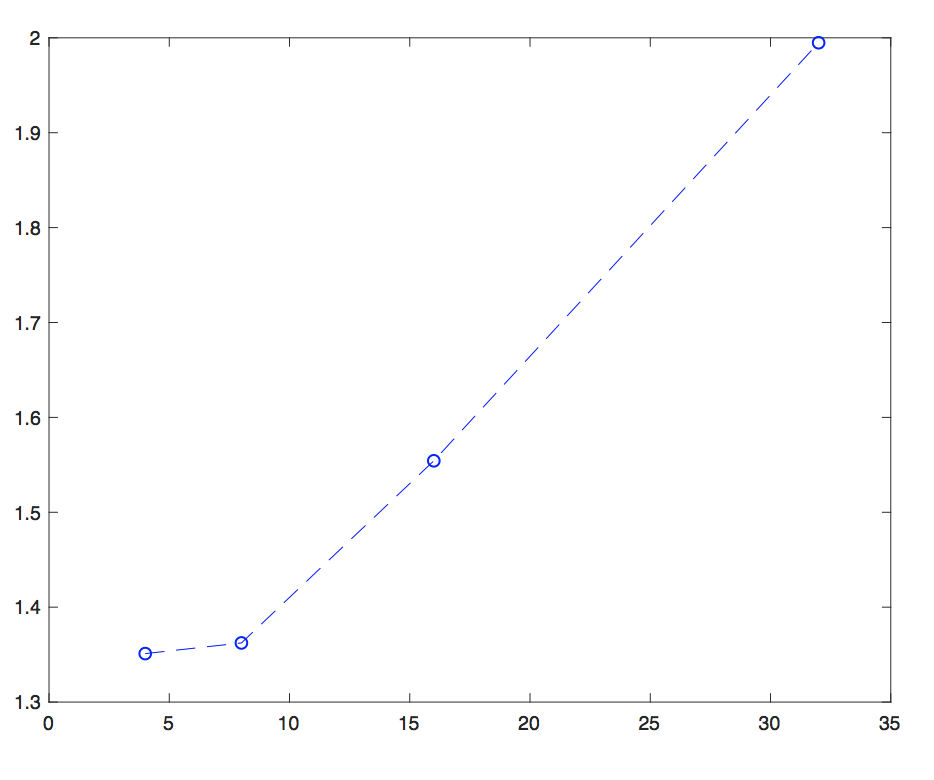}}

\caption{The reduced localization length $\Lambda_L=\xi_L/L$ as function of the transverse dimension $L$. We choose $\eta_{1,2}=\eta_{3,4}=1$, $\eta_{i,j}=0.1$ for other $(i,j)$ pairs. Upper panel: $\tan\alpha_{1,2}=1/6$ and $\tan\alpha_{3,4}=6$, the trend of localizing to a PH-Pfaffian state is clear. Lower panel: $\tan\alpha_{1,2}=2/3$ and $\tan\alpha_{3,4}=3/2$, a delocalized (metallic) state.}
\label{LambdaPH}
\end{figure}
\end{center}

\begin{center}
\begin{figure}

\captionsetup{justification=raggedright}

\adjustbox{trim={0\width} {0\height} {0\width} {0\height},clip}
{\includegraphics[width=1\columnwidth]{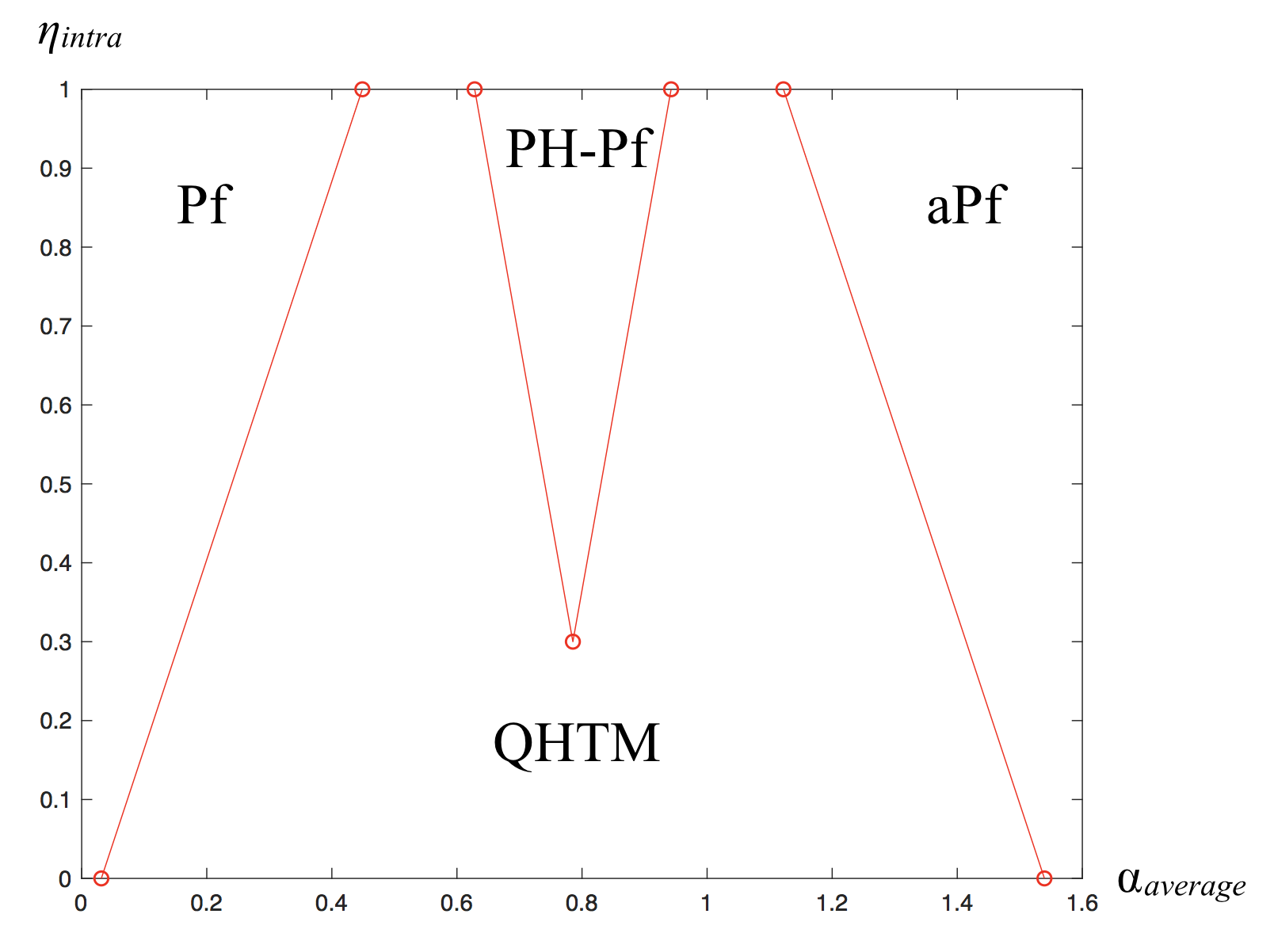}}

\caption{Phase diagram of the network model in Scenario B. We assume a weak $SO(4)$ flavor mixing with $\eta_{i,j}=0.1$, but a different mixing between flavor $(1,3)$ and $(2,4)$, given by $\eta_{intra}$ (vertical axis). The horizontal axis $\alpha_{average}$ describes the average particle-hole asymmetry. We assume $\Delta\alpha=\pi/6$, $\alpha_1=\alpha_3=max(\alpha_{average}-\Delta\alpha/2, \pi/100)$, and $\alpha_2=\alpha_4=min(\alpha_{average}+\Delta\alpha/2,\pi/2-\pi/100)$. We have also put in random vortices by having a random sign with equal probabilities on each link, which caused the metallic region in the small $\eta$ regime to significantly broaden.}
\label{PD2}
\end{figure}
\end{center}

We find that this could give rise to a localized intermediate state when $\alpha_{1,3}\ll \alpha_c$ and $\alpha_{2,4}\gg\alpha_c$, which is adiabatically connected to the PH-Pfaffian state. There is also a thermal metal phase separating the PH-Pfaffian from the Pfaffian and anti-Pfaffian states. See Fig.~\ref{LambdaPH} for examples of both. A crude phase diagram of the model is shown in Fig. \ref{PD2}. Notice that if the disorder strength in the real system is weak, the particle-hole asymmetry is switched suddenly as one tunes across $\nu_c$, so the entire intermediate region collapses to a first-order-like transition. The implied phase diagram is shown in Fig.~\ref{PD0}, assuming that as disorder strength grows, $\eta_{1,3}=\eta_{2,4}$ grow first, and the growth of other $\eta_{i,j}$'s comes much later due to the spatial separation between the modes.

Naively one may think that the PH-Pfaffian state is stable as long as the disorder is weak. Indeed if only weak random scattering is considered, i. e. $O_a$  is close to the identity matrix on every link ($\eta_{i,j}\ll 1$ for every $(i,j)$), then the system stays localized. However, we should include random vortices, i.e. random overall signs for each $O_a$, since they are not captured by weak scatterings (small $\eta_{i,j}$). We find that with random vortices (even at low density) and weak scattering the system actually delocalizes, which is the metallic region in Fig.~\ref{PD2} at small $\eta$.

\begin{center}
\begin{figure}

\captionsetup{justification=raggedright}

\adjustbox{trim={0\width} {0\height} {0\width} {0\height},clip}
{\includegraphics[width=1\columnwidth]{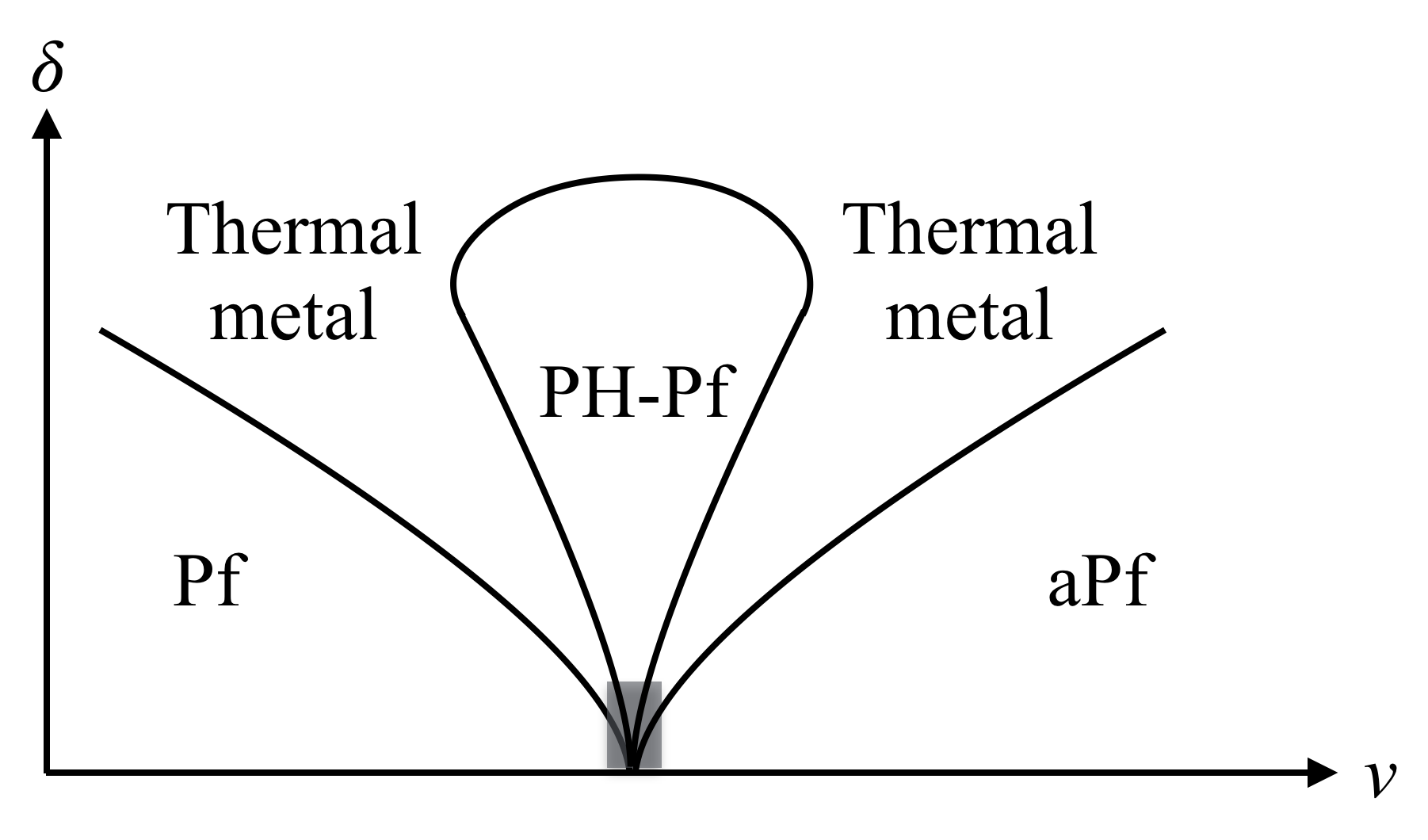}}

\caption{Schematic phase diagram of the network model with an approximate $U(1)$ symmetry. The shaded area represents a first-order-like transition in a finite sample.}
\label{PD2.1}
\end{figure}
\end{center}

In fact there is yet another possible mechanism to stabilize an intermediate PH-Pfaffian state, which we now briefly discuss. If there is a $U(1)$ symmetry that is approximately preserved at each link, namely a pair $(i,j)$ for which all the $O_a$'s approximately commute with $T_{i,j}$, then the Pfaffian and anti-Pfaffian states can be thought of as $\nu=1$ and $\nu=-1$ quantum Hall states with this conserved $U(1)$ symmetry. With the $U(1)$ symmetry the transition between the $\nu=1$ and $\nu=-1$ state will generically split into two transitions upon introducing disorder, with the symmetric point localized to a $\nu=0$ state, which in our case is nothing but the PH-Pfaffian state. Breaking the approximate $U(1)$ weakly will not destroy the localized $\nu=0$ state, but will broaden the transitions between $\nu=0$ to the other two phases into metallic regions. This scenario will give rise to a phase diagram that schematically look like Fig.~\ref{PD2.1}, which differs from Fig.~\ref{PD0} at weak disorder. However, for the physical system we do not see a good reason for the system to have an approximate $U(1)$ symmetry (it is important that the approximate symmetry is just $U(1)$ but not $U(1)\times U(1)$).

If we assume either of the above scenarios and interpret the observed $\kappa_{xy}=5/2$ as the disorder-induced PH-Pfaffian state, then we can predict the behavior of the system as one makes it less disordered. If the disorder strength is very low, the system will eventually land on either the Pfaffian or anti-Pfaffian state. But before that, one necessarily encounters a thermal metal phase.

\subsection{Scenario C: PH-Pfaffian and abelian phases}

We can consider an even more exotic scenario, in which all of the four domain wall modes are spatially separated from each other. This seems much harder to justify than even scenario B, but given how little we know about the domain wall structures, we will nevertheless discuss this scenario for completeness. For the network model, this scenario implies that (a) $\eta_{i,j}\ll1$ for all $(i,j)$ pairs, and (b) as one tunes $\nu$, the $\alpha_i$'s are tuned across $\alpha_c$ one by one.

It turns out that we need to consider two different possibilities even within scenario C, namely whether unpaired vortices are nucleated randomly within each Pfaffian/anti-Pfaffian domain. Since we expect some vortices to be nucleated within each domain, the question becomes whether it is cheaper to nucleate a single vortex (a $e/4$ quasi-particle) or a pair of it. It seems much more natural to assume that unpaired vortices can be nucleated by the disorder potential, giving rise to random $\pi$-vortices in the network model. But we will consider both possibilities here.

\subsubsection{Scenario C1: with random vortices}

In the presence of random vortices (even with relatively low density) we find that there is only a metallic phase in between the Pfaffian and anti-Pfaffian states. This gives qualitatively the same phase diagram as scenario A (Fig.~\ref{PD1}).

\subsubsection{Scenario C2: with only paired vortices} 

\begin{center}
\begin{figure}

\captionsetup{justification=raggedright}

\adjustbox{trim={0\width} {0\height} {0\width} {0\height},clip}
{\includegraphics[width=1\columnwidth]{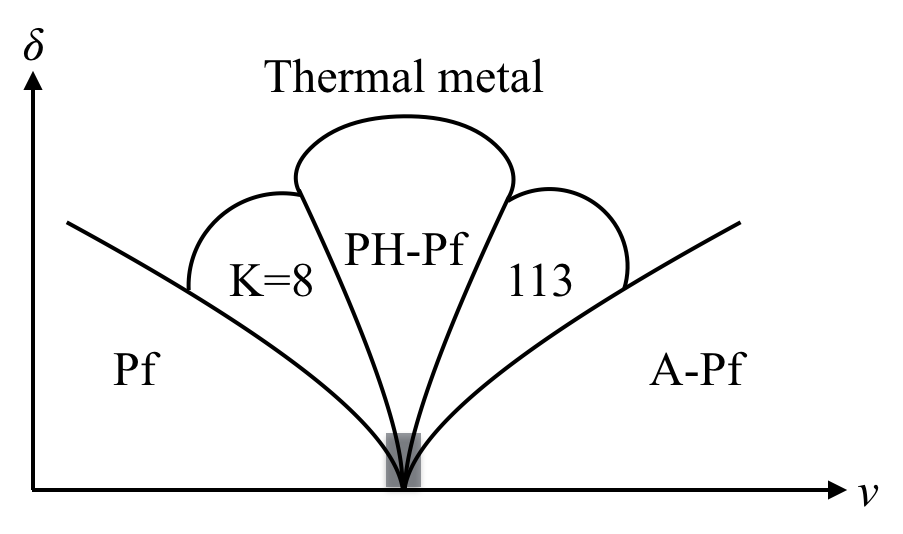}}

\caption{Schematic phase diagram of the network model in Scenario C2. The shaded area represents a first-order-like transition in a finite sample.}
\label{PD3}
\end{figure}
\end{center}

In this scenario there is no random $\pi$-flux in the network model, and we assume that flavor-mixing is always weak. As one tunes each $\alpha_i$ separately through $\alpha_c$, four separate direct transitions are observed. At each transition, $\kappa_{xy}/T$ jumps by $1/2$. The three intermediate phases are the PH-Pfaffian state, the abelian $K=8$ state equivalent to a $\nu=1/8$ Laughlin state of paired electrons\cite{Halperin1983} and its particle-hole conjugate known as the $(113)$ state\cite{YangFeldman}. A closely related transition between Pfaffian and $(331)$ phases has been discussed recently in Ref.~\cite{Yang}. A thermal metal state does not appear in this scenario as long as $\eta$ is sufficiently small. At very low disorder level, the four transitions collapses to a single first-order-like transition due to sudden switching of the particle-hole asymmetry. The schematic phase diagram is shown in Fig.~\ref{PD3}.

As one decreases the disorder strength gradually, one would eventually land on the anti-Pfaffian (or Pfaffian) state. The phase  diagram then implies that one would necessarily go through an intermediate abelian state with integer $\kappa_{xy}/T$. This is a very different prediction than what we reached from the scenario B.

\subsection{Some remarks on localization in class D}
\label{interpretation}

Now we offer some physical pictures to help us understand our numerical results, in particular how generic these results should be. 

We focus on the regimes where all the $\alpha_i$'s are sufficiently far away from $\alpha_c$ -- some much larger, and some much smaller. These are the regimes in which the system is expected to be gapped in the ``clean" limit (meaning $O_a=1$ for every link $a$). If the randomness is purely due to weak random scattering, so that $O_a$ is close to identity matrix on every link, then the gapped phase should be stable -- this is indeed what we found in such regimes. However, when random vortices are included, the system becomes easily delocalized -- the only exception is a PH-Pfaffian phase when the $(1,3)$ and $(2,4)$ scatterings become strong while the other scatterings remain weak. In fact, the drastic appearance of delocalization upon introducing random vortices was first noticed in Ref.~\cite{DMetal}, where a similar network model was studied, with only one chiral Majorana fermion propagating on each link. 

So why is the effect of random vortices so drastic? The physical reason is actually quite simple: in the absence of flavor mixing, a vortex ($\pi$-flux) creates a zero-energy mode for each flavor of Majorana fermions circulating around the vortex. A finite vortex density gives a finite density of zero modes. The average physical separation between two zero modes is given by the vortex density and is a fixed number, unlike in the strong Anderson localization scenario, where the separation between two levels with almost degenerate energy becomes arbitrarily large. Therefore it is not surprising that such systems will eventually delocalize and form thermal metals. We believe that this is the physical reason behind the metallic phases found in Ref.~\cite{DMetal}. Interestingly, this mechanism of delocalization by random $\pi$-flux can also be applied to the other two symmetry classes that allow metallic phases in $2D$ -- class AII (spin-orbit coupled insulators) and DIII (spin-orbit coupled superconductor).

Now we consider the effect of flavor mixing. We first consider the case with $\alpha_{1,3}\gg\alpha_c$ and $\alpha_{2,4}\ll\alpha_c$, which in the clean limit has Pfaffian islands circulated by fermions $\chi_{1,3}$ and anti-Pfaffian islands circulated by $\chi_{2,4}$, and overall gives the PH-Pfaffian state. The intra-island mixing $\eta_{1,3}$ ($\eta_{2,4}$) gives a splitting of the local energy levels on the Pfaffian (anti-Pfaffian) island, which is in analogy with the random chemical potential in the Anderson localization picture. We will thus schematically call this splitting $\Delta\mu$, even though this is really the energy level of some Bogoliubov fermions. This picture does not change even with random vortices, since a vortex creates an even number of zero modes on each island, which will be lifted by the random intra-island mixing. Therefore in order to have localization, we need strong intra-island mixing to produce a large variation of $\Delta\mu$, in accordance with the Anderson picture. In addition, we need to keep the inter-island mixing (e.g. $\eta_{1,3}$) to be small, otherwise an effective random hopping amplitude $\Delta t$ will be generated and will be comparable to $\Delta\mu$, which according to our numerical results leads to delocalization.

Now we consider the case with $\alpha_{1,2,3}\gg\alpha_c$ and $\alpha_{4}\ll\alpha_c$, which in the clean limit gives the abelian $K=8$ state (or its particle-hole conjugate). It is now easy to see why it is hard to avoid delocalization from random vortices: each island has an odd number (one or three) of chiral Majorana fermions circulating it, and in the presence of a $\pi$-vortex, an odd number of Majorana zero modes are generated, which cannot be completely lifted through intra-island coupling -- generically one zero mode will survive. 

Notice that this is not to say that the abelian $K=8$ phase is absolutely impossible to appear in the presence of random vortices. But in order for it to appear, the Majorana fermions on the domain walls must further mix with some other degrees of freedom. For example if disorder is sufficiently strong, the domain wall fermions can mix with some gapped degrees of freedom, which could then lead to localization. Of course at strong disorder many other phenomena could take place and it is not clear whether $K=8$ can really be stabilized there. Another possibility is that the domains are quite small, so the zero modes from the domain walls can mix with the zero modes associated with the $\pm e/4$ quasi-particles (the vortices) inside the domains. In reality we expect the domains to be at least of the same scale as the typical wavelength of the disorder potential, which should be much larger than the correlation length of the Pfaffian/anti-Pfaffian state. So this ``small domain wall" mechanism is unlikely to apply. Therefore we conclude that in real systems the abelian $K=8$ phase and its particle-hole conjugate are very unlikely to appear. Instead there is the metallic phase that sits between PH-Pfaffian (if it exists) and Pfaffian/anti-Pfaffian as one tunes a parameter controlling the degree of particle-hole asymmetry, say by tuning the filling fraction if the Hall plateau does not disappear.

Finally, we turn to  the ``improper vortices", introduced at the end of Sec.~\ref{networksection}, but not included in any of the calculations discussed above. By now it is obvious why the improper vortex, seen by an odd number of fermions around a domain, always leads to delocalization:  it induces an odd number of Majorana zero modes on the domain wall, which cannot be lifted locally, and eventually proliferates and produces a metal.

\section {Additional Remarks}

One may  ask whether disorder can help stabilizing the PH-Pfaffian state through some  mechanism other than the ones considered above. For example, in the Son-Dirac picture the PH-Pfaffian state is obtained through an $s$-wave pairing between Dirac composite fermions, while the Pfaffian and anti-Pfaffian phases have $d\pm id$ pairing. It is then natural to ask whether disorder would suppress $d$-wave pairing more than $s$-wave through some Anderson-like mechanism\cite{AndersonTheorem}. However, since the disorder potential locally breaks particle-hole symmetry, the Dirac composite fermions effectively see a disorder potential which locally breaks time-reversal symmetry. So the usual Anderson theorem does not apply in this case. A direct calculation shows that the disorder-suppression of the $s$-wave pairing is at least as much as that of the $d$-wave pairing. So stabilizing the PH-Pfaffian phase through Anderson-like mechanism does not seem to be a viable approach.

We also note that  though it requires certain nontrivial assumptions to stabilize PH-Pfaffian state in our network model, it would be even much harder to stabilize the PH-Pfaffian phase had the clean limit been the abelian $K=8$ phase (or its particle-hole conjugate). This is because a mixture of $K=8$ phase and its PH-conjugate will have two (instead of four) chiral Majorana modes propagating on each domain wall, and a PH-Pfaffian phase requires one mode to be left-turning and the other one to be right-turning. Following our discussion in Sec.~\ref{interpretation}, once we introduce random vortices into the domains, even with relatively small density, the system becomes delocalized. From this point of view the existence of a $K=5/2$ plateau can be seen as a supporting evidence that the system in the clean limit should be a non-abelian phase.

We have discussed the localization/delocalization of the disordered Majorana fermions in a free fermion framework (although the parameters of the network model itself may be determined by domain wall energetics). In the real system there are certainly some residual interactions among the Majorana fermions, and it is important to know whether the interactions change the results qualitatively. The localized phases, including the PH-Pfaffian state, are expected to be stable against weak interactions. For the quantized Hall thermal metal (QHTM) phase, it was argued in Ref.~\cite{Jengetal}, based on the replica non-linear sigma model, that weak interactions do not change the behaviors of the class-D thermal metal. Therefore the general picture we obtained is unaffected by the residual interactions as long as they are not too strong.

Although we have focused here on the possible effects of disorder-induced inhomogeneity, one might also consider the possibility of important effects due to systematic sources of inhomogeneity.  For example, if the Pf state were to be favored by conditions at the sample boundary, while the aPf state was favored in the bulk due to Landau-level mixing, there would then be a domain wall  separating the two phases, in the vicinity of the boundary.  If this domain wall were sufficiently thick, so that it engenders a spatial separation between the two inner Majorana modes and the two outer Majorana modes propagating along the wall, it is conceivable that the outer modes would reach thermal equilibrium with the sample edge while the inner modes remain isolated.  In this case, the sample would show a thermal conductance with $K=5/2$.

 \section {Conclusions}
 
 In this paper we began with a microscopic analysis of the disordered $\nu=5/2$ state, assuming that in the clean limit,  only  the Pfaffian or anti-Pfaffian states are realized. This led us to construct an effective network model representing the edge states surrounding puddles of these phases nucleated by disorder. The effective parameters of this network model should be determined by the underlying energetics of domain walls, which evolve on changing the filling.  Assuming that the Pfaffian or anti-Pfaffian will tend to be favored away from the center of the quantized Hall plateau, 
 we identified three scenarios for this evolution. First, the system could switch in a `first order' type fashion, between these two states, which is expected to occur if disorder is relatively weak and domain walls are expensive. In this scenario $K$ rapidly switches between $7/2$ and $3/2$ on sweeping $\delta \nu$. Next, if disorder is stronger, we expect a thermal metal phase to intervene. Despite conducting heat like a metal, this Quantized Hall Thermal Metal is an electrical insulator with Hall conductance quantized to $\sigma_{xy}=\frac{5}{2}\frac{e^2}{h}$, and it represents a new state of matter  induced by disorder. Finally, we discussed conditions under which an intervening PH-Pfaffian topological order is stabilized, which would be consistent with the current experimental observations. Although for generic parameters our network model does not favor this behavior, this scenario may be realized with certain  additional constraints that either (i) effectively  assume that the PH-Pfaffian is a subdominant phase nucleated in the domain walls or (ii) there is an effective U(1) symmetry for Majorana modes along domain walls, that constrain the scattering events.  In view of the difficulty we have encountered in finding realistic parameters that lead to stabilization of the PH-Pfaffian by disorder, we suggest that there is a need for further numerical calculations  to see if there might be some range of interaction parameters for which the PH-Pfaffian can actually emerge as the ground state in a system without disorder.

Finally we close with some comments regarding future directions.  The picture of disorder induced nucleation of domains of Pfaffian and anti-Pfaffian states can be checked in numerical simulations. In particular a key parameter, the energy splitting between the $e/4$ quasi-electron versus the quasi-hole which is expected to drive the nucleation, can be calculated in the clean limit. We note that there are other experimental platforms where an incompressible state has been obtained in the half filled Landau level. In particular, a robust plateau was observed in Bernal bilayer graphene \cite{Zibrov2017}, where the effect of Landau level mixing is expected to be stronger than in the case of GaAs quantum. If so, the delicate balance between the Pfaffian/antiPfaffian states invoked here would be absent. Finally, on a more general note, the picture of disorder inducing entirely new topological phases that are not present in the clean limit is an intriguing possibility that, if confirmed, would have important consequences for other quantum Hall states as well as more generally for correlated quantum systems.

Note added: during the completion of this work, we became aware of the work by Mross, et. al\cite{Mrossetal}, which also addressed the possibility of a disorder-induced PH-Pfaffian state at $\nu=5/2$. The model used in their analysis differs in some respects from ours, and there are differences in our conclusions about the most likely phase diagrams.
 
{\textbf{Acknowledgments}: We thank M. Barkeshli, D. Feldman, Y. C. He, R. H. Morf, T. Senthil, A. Stern and M. Zaletel for helpful discussions. CW thanks N. Wang for advice on programming. We thank D. Mross and A. Stern for sharing their manuscript prior to publication. CW was supported by Harvard Society of Fellows. AV was supported by a Simons Investigator grant.  BIH acknowledges support from Microsoft Station Q.

%\appendix

\bibliography{PH-5_2}

\end{document}